\documentclass[fleqn,10pt]{wlscirep}
\usepackage[utf8]{inputenc}
\usepackage[T1]{fontenc}
\title{Coloured Digital Terahertz Holography Within 1.39 -- 4.25 THz range }
\usepackage{amsmath} % for \longrightarrow
\author[1,*]{Rusnė Ivaškevičiūtė-Povilauskienė}
\author[1]{Linas Minkevičius}
\author[1]{Ignas~Grigelionis}
\author[2]{Agnieszka~Siemion}
\author[1]{Domas Jokubauskis}
\author[3]{Kęstutis Ikamas}
\author[3]{Alvydas Lisauskas}
\author[1]{Gintaras Valušis}
\affil[1]{Center for Physical Sciences and Technology, Department of Optoelectronics, Vilnius, LT-10257, Lithuania}
\affil[2]{Warsaw University of Technology, Faculty of Physics, Warsaw, 00-662, Poland}
\affil[3]{Vilnius University, Institute of Applied Electrodynamics and Telecommunications, Saulėtekio av. 3, Vilnius 10257, Lithuania}

\affil[*]{rusne.ivaskeviciute@ftmc.lt}

\begin{abstract}
Terahertz coloured digital holography ranging from 1.39 THz up to 4.25 THz is demonstrated. It is shown
that it can be applied for the investigation of low-absorbing objects, and it is illustrated via inspection of stacked graphene layers placed on high-resistivity silicon substrate. Holograms are recorded using an optically-pumped molecular THz laser operating at discrete emission lines of 1.39~THz, 2.52~THz, 3.11~THz, and 4.25~THz frequencies and it unveiled that phase-shifting methods allow to qualitatively reconstruct coloured THz holograms with improved quality achieved by removing unwanted information related with so-called DC term and conjugated beam forming virtual image.
\end{abstract}
\begin{document}

\flushbottom
\maketitle
\thispagestyle{empty}

%\noindent Please note: Abbreviations should be introduced at the first mention in the main text – no abbreviations lists. The suggested structure of the main text (not %enforced) is provided below.

\section*{Introduction}
The ability of terahertz (THz) radiation to penetrate into many different non-conducting materials that are opaque for visible light and that exhibit here very low absorption or %even 
uniform transparency offers a unique possibility for THz imaging to look into and through different objects \cite{Mittleman2018, Valusis2021, Castro-Camus2022}. 
When item under test exhibits low absorption, a coherent recording techniques like heterodyne 
\cite{Glaab2010, Minkevicius2011, Yuan2019} or homodyne \cite{Jokubauskis2019, Ikamas2021} detection gain their distinct importance as it allows to discriminate the content or monitor the quality of materials via their induced phase shifts. A particular role in the family of coherent imaging techniques can be attributed to THz holography due to its fascinating capability to reconstruct 3D images and  extend hence the borders of practical uses into the possible observation of subwavelength-size defects or nonuniformities in inspected materials or investigated objects \cite{Heimbeck2020}.

Although digital holography \cite{Javidi2021} is well established at optical wavelengths \cite{Sheridan2020, Nehmetallah12,goodman1967digital,kreis1997suppression}, its implementation in the THz range still remains a challenging issue due to the still remaining issue of low power of THz radiation sources and reliable sensitive detectors operating at room temperature. Furthermore, it deserves noting another very important concern -- it is the relation between the wavelength, the size of the optical elements and the scanned area, which for THz range is very unfavourable. Optical setups constructed for THz range of radiation suffer from large diffraction effects and undesirable influence of limited apertures of optical elements, which are tremendously more significant if compared to a visible range. The lack of reasonable size of detector arrays with relevant pixel pitch and large enough elements diameters resulted in point-to-point imaging -- focusing the light on the sample and then on the detector with subsequent shifting of the sample. Such configuration enables using only the radiation propagating around the optical axis which diminishes unwanted diffraction effects. It must be underlined that plane-to-plane imaging requires much larger apertures of all used optical elements. In case of registering THz holograms, both approaches are possible and additional methods (like phase-shifting techniques \cite{Siemion2021}, filtering in Fourier domain \cite{siemion2021spatial,liu2021terahertz} or many others) of removing unwanted components from reconstruction are required.

Up to now, it was a large variety of interesting efforts and different approaches dedicated to the development of THz holography \cite{holo2012} such as in-line holography \cite{Rong2015, hartmut-holo2019}, off-axis holography \cite{Locatelli2015, Zolliker2015, Valzania2017},
near-field holography \cite{Amineh2011} and time-reversal holographic imaging of hidden objects \cite{Wu2015}.  Innovations in THz holography via step phase-shifting methods allow for improvement in the quality of the recorded holographic images \cite{Sun2013, zuo2018, Siemion2021}. Very recently, lensless Fourier-transform THz digital holography for full-field reflective imaging was demonstrated using an optically-pumped molecular THz laser operating at 2.52~THz \cite{Zhang2022}. 
It is worth noting that holographic images, as a rule, are recorded using one wavelength obtaining thus monochromatic images. 

In the given work, we enrich the family of THz holographic experiments by demonstrating coloured digital holography ranging from 1.39~THz up to 4.25~THz and unveil its possible application for the low-absorbing objects via inspection also of stacked graphene layers. Relying on the setup based on an optically-pumped molecular THz laser operating at discrete emission lines of 1.39~THz, 2.52~THz, 3.11~THz, and 4.25~THz frequencies as well as THz sensitivity of broadband nanometric field effect transistors \cite{zdanevicius:2018} we demonstrate that phase-shifting (PS) methods \cite{Siemion2021} allow to qualitatively reconstruct coloured THz holograms with improved quality especially of phase retrieval achieved by removing unwanted information related with so-called DC term and conjugated beam forming virtual image. Phase values corresponding to the phase shifts introduced by the object were retrieved at different wavelengths providing thus a way for deeper inspection of the investigated materials.

\subsection*{Results and Discussion}

Holograms are recorded by employing the setup depicted in Fig.~1. The coherent THz radiation delivered by optically-pumped molecular laser (FIRL-100, Edinburgh Instruments Ltd) was focused using parabolic mirror with \textit{f}=5~cm ({PM$_1$}) and, after chopper modulation, was collimated by PM$_2$. The collimated THz beam reflected by a flat gold-coated mirror (M$_1$) was then divided into two parts by the 525 \textmu m-thick high resistivity silicon beam splitter (BS$_1$). One part was focused on the pyroelectric detector (D$_1$) by the third parabolic mirror (PM$_3$) and served as the laser power reference. The reference beam data was later used to compensate laser mode instabilities in the imaging data. The second part of the beam was divided into two arms once again by the second beam splitter (BS$_2$, equivalent to BS$_1$). The beam in the first arm was reflected from the second flat mirror (M$_2$), passed through the sample and, subsequently, after passing through the third beam splitter (BS$_3$), reached the second detector as the object beam. 
The beam in the second arm also was reflected from the third flat mirror (M$_3$), and its phase was shifted by the introduced phase delay element PS. This beam served as a reference beam. Subsequently, after reflecting from BS$_3$ it interfered with the beam from the first arm and also reached the D$_2$. The transmitted beam power was registered using nanometric field-effect transistor-based terahertz detectors (TeraFETs) with integrated patch antennas. The devices have been fabricated using a standard 65-nm CMOS technology, their operation and design are in detail described elsewhere \cite{zdanevicius:2018}. Since measurements using room-temperature CMOS TeraFET do not requires any modulation, however, for the detector response measurement, a lock-in technique was used with the mechanical chopper "C" to increase the signal-to-noise ratio in the measurement system.

\begin{figure}[ht!]
    \centering
    \includegraphics[width=1\columnwidth]{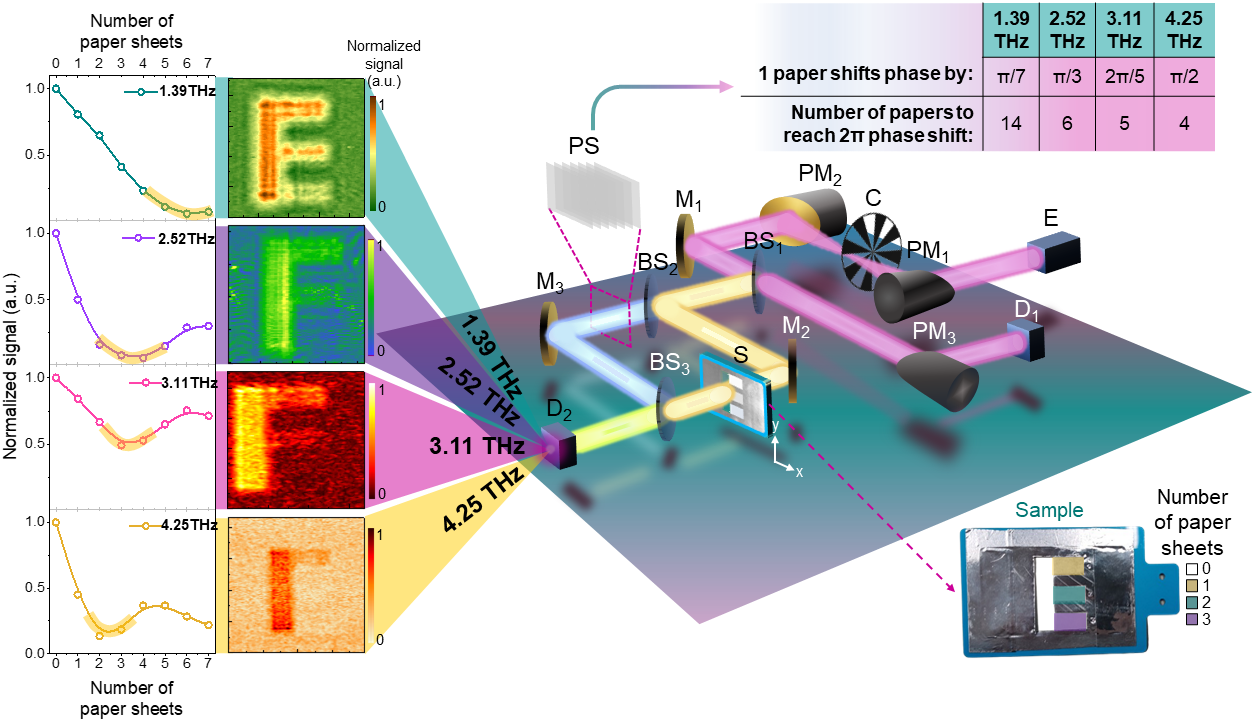}
    \caption{Optical setup for the recording THz digital holograms at 1.39~THz, 2.52~THz, 3.11~THz, and 4.25~THz. Letter E in the scheme of the optical system denotes THz source -- optically-pumped molecular THz laser; PM$_1$, PM$_2$, and PM$_3$ are parabolic mirrors; C is a chopper; M$_1$, M$_2$, and M$_3$ label gold-coated flat mirrors; BS$_1$, BS$_2$, and BS$_3$ mark high resistivity silicon beam splitters; S is the imaged sample; thin napkins, referred to in the rest of this article as paper sheets served as a simple phase shifter; D$_1$ and D$_2$ are detectors. A signal intensity dependence on the number of paper sheets, added to a PS element is depicted on the left side for four frequencies. A transparent yellow line marks the minima of the normalised signal. The example of the registered holograms at four different THz frequencies is given in the left coupled panel. A table in the top right corner presents the corresponding PS induced by one paper sheet for each frequency. Inset at the bottom right corner depicts a sample made from aluminum foil with a cut-out "E" letter shape. Parts of the sample are covered by different numbers of paper sheets to produce different PS. Here, it is indicated by the different colours.}
    \label{fig:setup}
\end{figure}

The investigated sample (marked as "S" in Fig.~\ref{fig:setup} ) was raster scanned with a spatial resolution of 0.5~mm in \textit{x} and \textit{y} directions. A special sample dedicated to illustrate the holographic images was fabricated from metallic foil of 55x85~mm$^{2}$ in dimensions with an "E" letter-shaped aperture (shown at the bottom of the right corner of Fig.~\ref{fig:setup}). The vertical part (marked in white color) was completely hollow. Meanwhile, the horizontal parts were covered with 1 (marked as yellow), 2 (marked as blue) and 3 (marked as purple) paper sheets (thin napkins) that shift the phase (0.2$\pi$; 0.36$\pi$; 0.41$\pi$ and 0.55$\pi$ for the frequency 1.39~THz, 2.52~THz, 3.11~THz and 4.25~THz, respectively, according to time-domain spectroscopy measurements and determined particular  thickness of 0.12~mm and refractive index values -- 1.177; 1.175; 1.165 and 1.160, respectively). Due to the fact, that THz time-domain spectroscopy technique has decreasing efficiency (thus signal-to-noise ratio) for higher and higher frequencies, the additional measurements were conducted. Moreover, the refractive index value calculation is strongly dependent on the thickness of the sample - which in case of napkin tissue might be small and introduce large uncertainty. A signal intensity dependence on the number of paper sheets was calculated and a transparent yellow line in Fig.~\ref{fig:setup} marks the minima of the normalised signal - the corresponding values are: 0.14$\pi$; 0.33$\pi$; 0.4 $\pi$ and 0.5 $\pi$, respectively for each frequency.

As one can see, as the frequency increases, the PS also increases, meaning that a few paper sheets are needed to achieve a complete 2$\pi$ PS. For example, at 1.39~THz $2\pi$ PS can be accumulated using 14 paper sheets, at 2.52~THz the number of paper sheets decreases to 6, at 3.11~THz -- 5 paper sheets are needed and, finally, at 4.25~THz only 4 paper sheets are required. These estimates are given in the table on the top right corner of Fig.~\ref{fig:setup}. The left panel demonstrates shift of the local signal minima on variation of number of paper sheets at different frequencies. 

A simple PS element fabricated of a different number of the paper sheets (thin napkins) was placed in the second arm of the interferometer (reference beam) and was used to induce a PS -- as one paper sheet corresponds to 0.2$\pi$; 0.36$\pi$; 0.41$\pi$ and 0.55$\pi$ (or 0.14$\pi$; 0.33$\pi$; 0.4 $\pi$ and 0.5 $\pi$, according to the experimental evaluation the signal intensity dependence on the number of paper sheets), for the frequency 1.39~THz, 2.52~THz, 3.11~THz and 4.25~THz, respectively.

\begin{figure}[ht!]
    \centering
    \includegraphics[width=0.75\columnwidth]{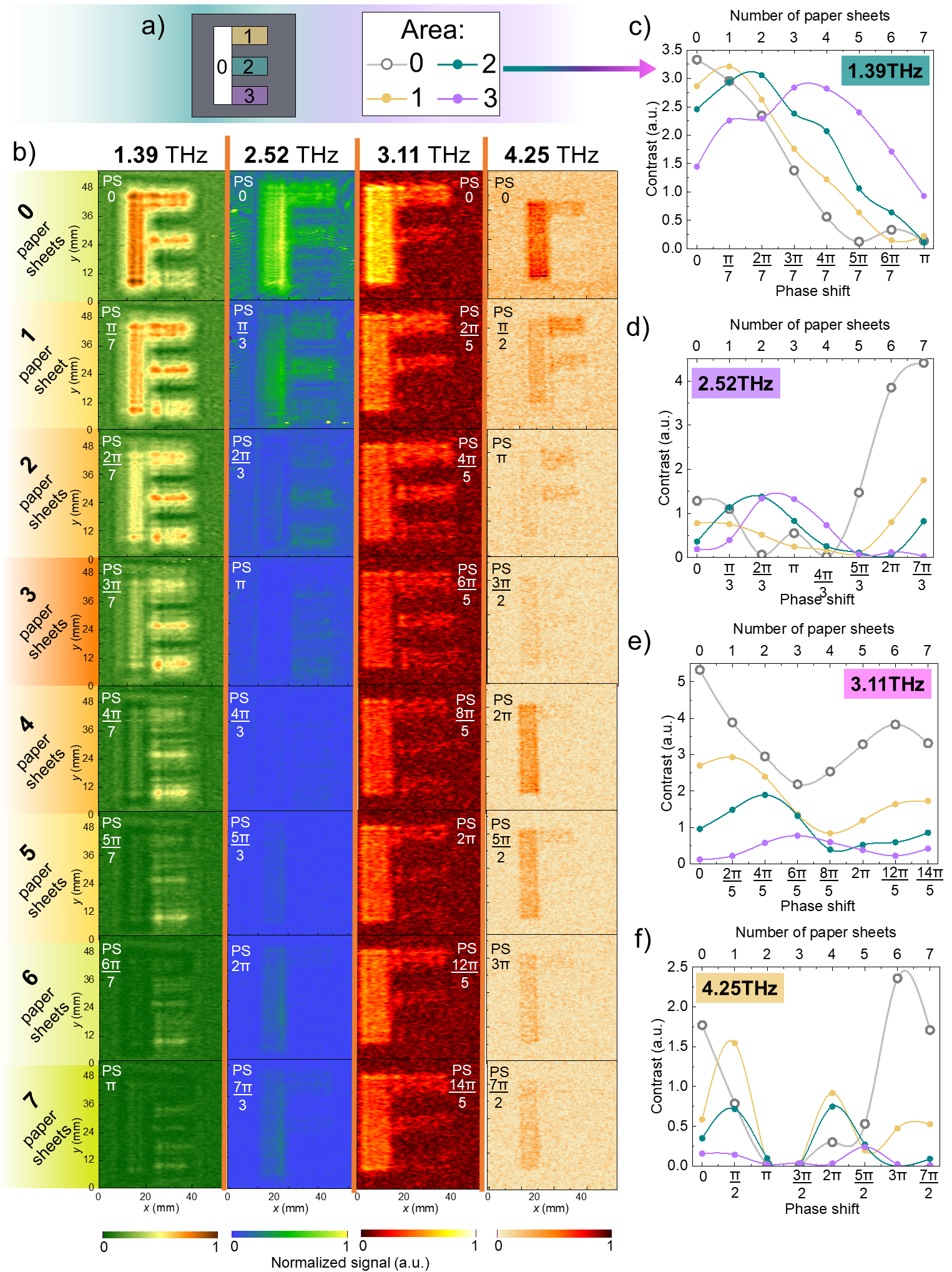}
    \caption{Experimentally recorded intensity distribution in E letter-shaped sample at different frequencies varying the PS. 
    {\bf Panel (a):} the scheme of the imaged sample consisting of different areas covered with varying number of paper sheets. Gray area is a metal foil, white area marks hollow part, yellow area is covered with 1 paper sheet, green area with 2 and purple area with 3 paper sheets. 
    {\bf Panel (b):} Experimentally obtained two-dimensional intensity distributions at four frequencies: 1.39~THz, 2.52~THz, 3.11~THz, and 4.25~THz. Image was recorded 7 times placing an additional paper sheet to the reference beam to introduce a corresponding PS. Coloured scale is normalized to the maximum signal. 
    {\bf Panels (c-f):} Contrast dependencies on the number of paper sheets placed in the reference beam (top scale) or corresponding PS (bottom scale) at four different frequencies: 1.39~THz (c), 2.52~THz (d), 3.11~THz (e) and 4.25~THz (f). In each case, grey, yellow, green and purple coloured lines with symbols represent the contrast values at corresponding areas of the sample marked with the same colours. Shifting the phase in the reference beam by $\pi$, the contrast decreases by 96\% for the 1.39~THz frequency, by 57\% for 2.52~THz, by 59\% for 3.11~THz and by 98\% for 4.25~THz.}
    \label{fig:E_experiment}
\end{figure}

Corresponding intensity distributions in holograms are presented in the left coupled panel in Fig. \ref{fig:setup} and for all PS in reference beam in Fig. \ref{fig:E_experiment}. It is worth noting that during these measurements, no additional PS was placed in the object beam, and the PS was introduced only by the sample parts that were covered with 1, 2 and 3 paper sheets. 
As it is seen, at 1.39~THz frequency, all four different areas of the imaged sample are clearly visible.

When the frequency increases to 2.52~THz and 3.11~THz, only three areas the sample are visible in the hologram, while at 4.25~THz the intensity distribution in the hologram becomes less pronounced as due to increasing attenuation of the THz radiation intensity with the increase of the number of PS paper sheets.

The experiment was repeated for each frequency, with an increasing PS achieved by adding an extra sheet of paper in the reference beam. The two-dimensional signal intensity distributions are shown in Fig.~\ref{fig:E_experiment}(b). These results are in a good agreement with the already discussed dependence of the signal intensity on the PS. It is also visible that for the consecutive holograms the attenuation introduced by the paper PS is increasing with a number of paper sheets and with recording frequency as well.

As it is seen, "E" letter shaped sample is well-resolved in all recorded holograms except 4.25~THz. Varying the PS, its different parts becomes more pronounced. When the PS reaches 2$\pi$, hologram distributions look similar to that without a PS, but with the strongly reduced intensity due to introduction of paper sheets which except PS introduce also the attenuation for higher frequencies. This effect correlates with the intensity and its dependence on the PS given in 
Fig.~\ref{fig:E_experiment}(c-f) panels. It should be also underlined that decreasing intensity of the reference beam results in larger difference of intensities of interfering beams, thus the lower contrast of recorded hologram.

Moreover, the contrast values were estimated for each separate area of the imaged sample in the hologram and results are represented in Fig.~\ref{fig:E_experiment}(c-f). The contrast was determined by $C$=($I_a$-$I_n$)/$I_n$, where $I_a$ is the average signal intensity in the specified area (marked with 0, 1, 2 or 3) and $I_n$ is the noise or average intensity of the metal covered area. As one can see, at 1.39~THz, 3.11~THz and 4.25~THz frequencies the contrast graphs of 0 area coincides with graphs in Fig.~\ref{fig:setup}. Although, this does not correspond to the 2.52~THz frequency, where very weak signal, close to the noise floor, is detected.
%and maybe noise is higher - here we use Weber contrast - 
%maybe calculate Michelson contrast - $(I_max-I_min)/(I_max+I_min)$?}. 
For 1.39~THz, as it was indicated previously, with additional 7 paper sheets only small PS is observed. For this reason contrast graphs of 1,2 and 3 areas lie quite close to the 0 area graph. 
For each frequency all the sample areas exhibit the same tendency.

\begin{figure}[ht!]
    \centering
    \includegraphics[width=0.7\columnwidth]{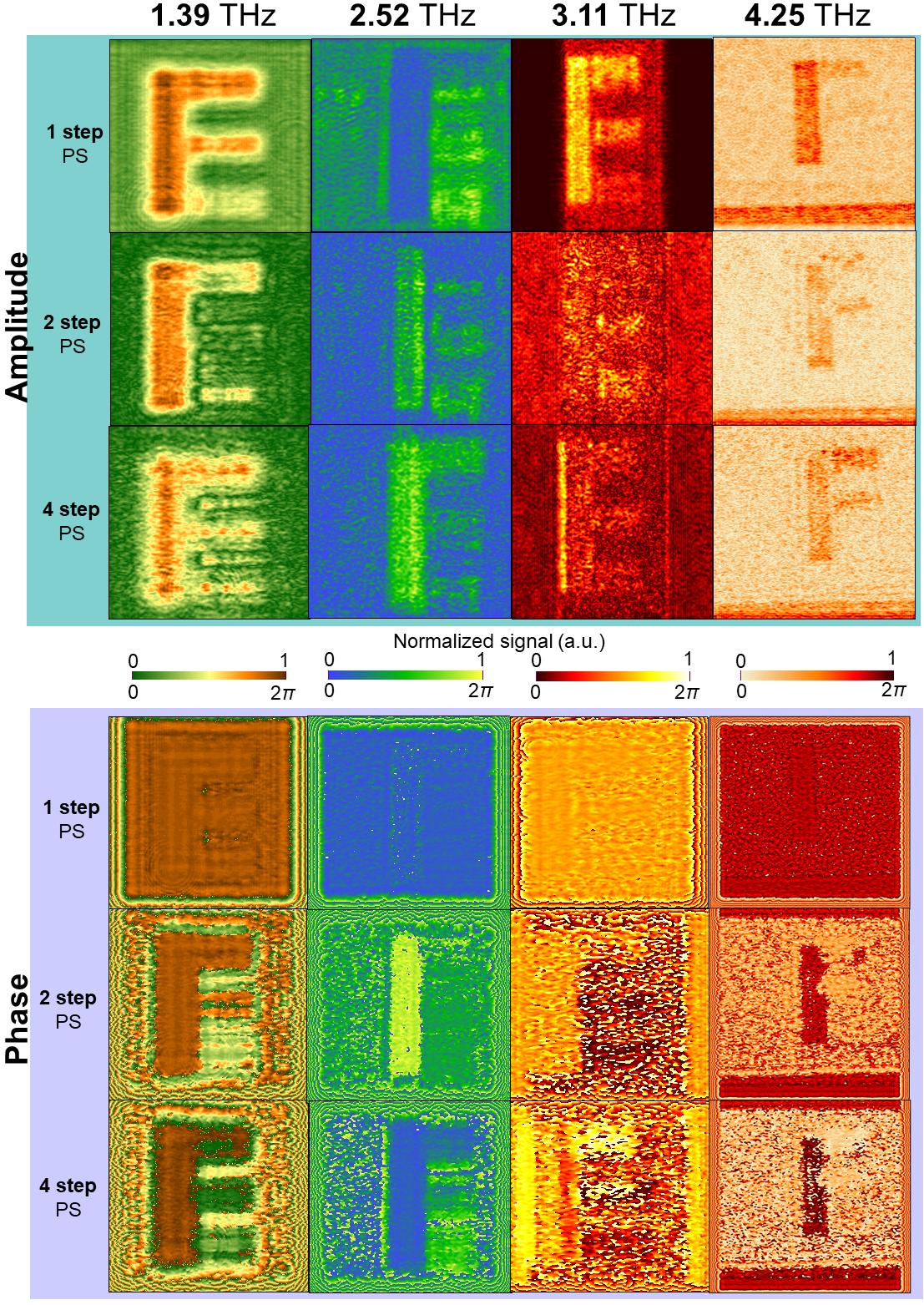}
    \caption{Amplitude (top part) and phase (bottom part) distributions of the reconstructed holograms at four different frequencies: 1.39~THz, 2.52~THz, 3.11~THz and 4.25~THz. In each case the reconstructions in the first row are obtained using a single experimentally measured hologram. The reconstructions in the second row are obtained using 2-step PS method, while the reconstructions in the third row are obtained using 4-step PS method. Coloured scales are normalized to the maximum values and are different for each frequency. The amplitude distributions of obtained images is similar in all cases, however, the phase distributions differ a lot. Due to the removal of unwanted terms in the reconstructed light field distribution -- the PS introduced by different areas of "E" letter starts to be visible and distinguishable. Note that only for 4-step PS method all areas have different phase level in reconstructed image, especially for 1.39~THz and 2.52~THz. In case of 3.11~THz and 4.25~THz reconstructed phase distributions different areas are less distinguishable, but still are the most different for 4-step PS method.}
    \label{fig:Reconstructions}
\end{figure}

After recording the holograms in the detector plane, each of them was reconstructed numerically. The 4-step PS algorithm was used to reconstruct the information only about the recorded object \cite{zuo2018,siemion2010one}. For comparison, the reconstructions were also conducted for 2-step PS algorithm \cite{siemion2021spatial} and the single hologram reconstructions. The modified convolution approach \cite{sypek1995light} was used to propagate the light field distribution to the object plane. Each hologram was scanned with 0.5~mm resolution having the total size of 54 x 54 mm. 
For the reconstruction the whole scanned hologram area was placed inside 1024 x1024 pixel matrix which was rescaled 4~times to increase the reconstructed image resolution. The digital reconstruction was performed on 4096 x4096 pixels calculation matrix with sampling of 0.125~mm. 
The registered intensity pattern was recalculated according to the desired algorithm (4-step, 2-step PS or single hologram reconstructions) and propagated back at a distance of 95~mm corresponding to the distance between the object and the hologram planes. The propagation was conducted using off-axis approach which does not have any approximations. From each intensity pattern of the hologram, amplitude and phase distributions in image plane obtained and for all frequencies and algorithms are shown in Fig.~ \ref{fig:Reconstructions}. 

Before proceeding further, it is worth to remind that the sample was a transparent "E" letter which parts were covered with the consecutive number of paper sheets. They introduce particular PS, and the paper tissue itself introduced also some additional attenuation, especially for higher frequencies. Thus, in the reconstructed images of our object one should expect relatively uniform amplitude pattern -- dependent only on influence of attenuation of the paper sheets, and phase values corresponding to the different PS introduced by the sample.

\begin{figure}[ht!]
    \centering
    \includegraphics[width=0.7\columnwidth]{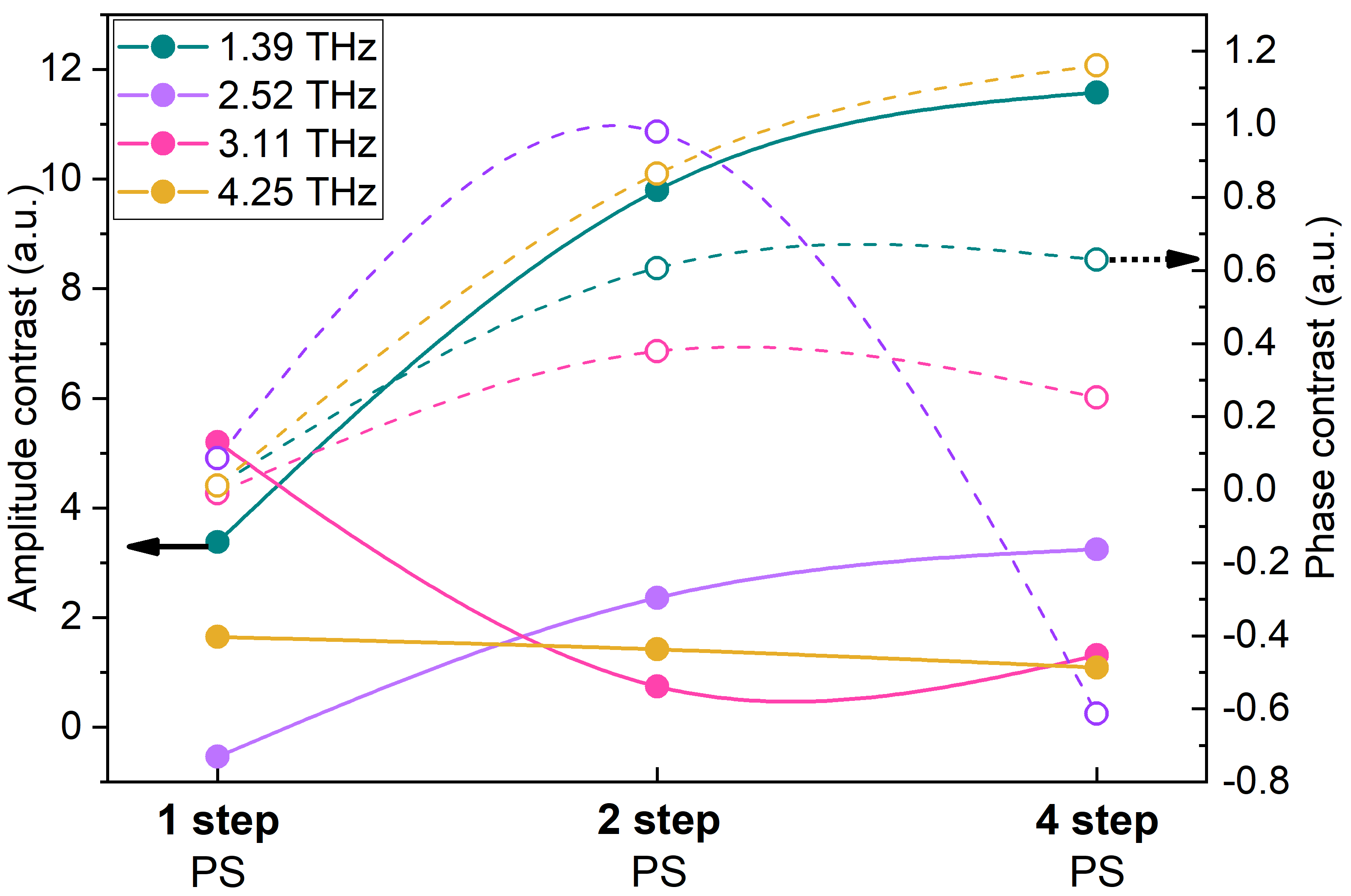}
    \caption{Contrast of reconstructed amplitude and phase for sample "E" at different frequencies: 1.39~THz, 2.52~THz, 3.11~THz and 4.25~THz.  Reconstructions were obtained from a single (1-step) hologram and using 2-step and 4-step PS techniques. Contrast was obtained from the area "0" without paper sheets. Results of amplitude contrast are represented in full circles and straight lines, while the phase contrast results are marked as open circles connected with dashed lines.} 
    \label{fig:Contrast_Reconstructions}
\end{figure}

For comparison, contrast of amplitude and phase 
reconstructions for holograms recorded in all frequencies  when the reconstructions were obtained from a single (1-step) hologram and using 2-step and 4-step PS
techniques, for are presented in Fig.~ \ref{fig:Contrast_Reconstructions}. As one can see, amplitude contrast is well pronounced for 1.39~THz, and diminishes with frequency for all techniques used. Application of 2-step PS and 4-step PS allow to remove additional interferences and make the amplitude more uniform, but do not totally remove interferences from some parts of the setup. Regarding phase contrast, 4 step PS technique gives the best results. Nonmonotonical phase contrast behaviour at 2.54~THz can be associated probably with unwanted interference effects in the setup.

\begin{figure}[ht!]
    \centering
    \includegraphics[width=1\columnwidth]{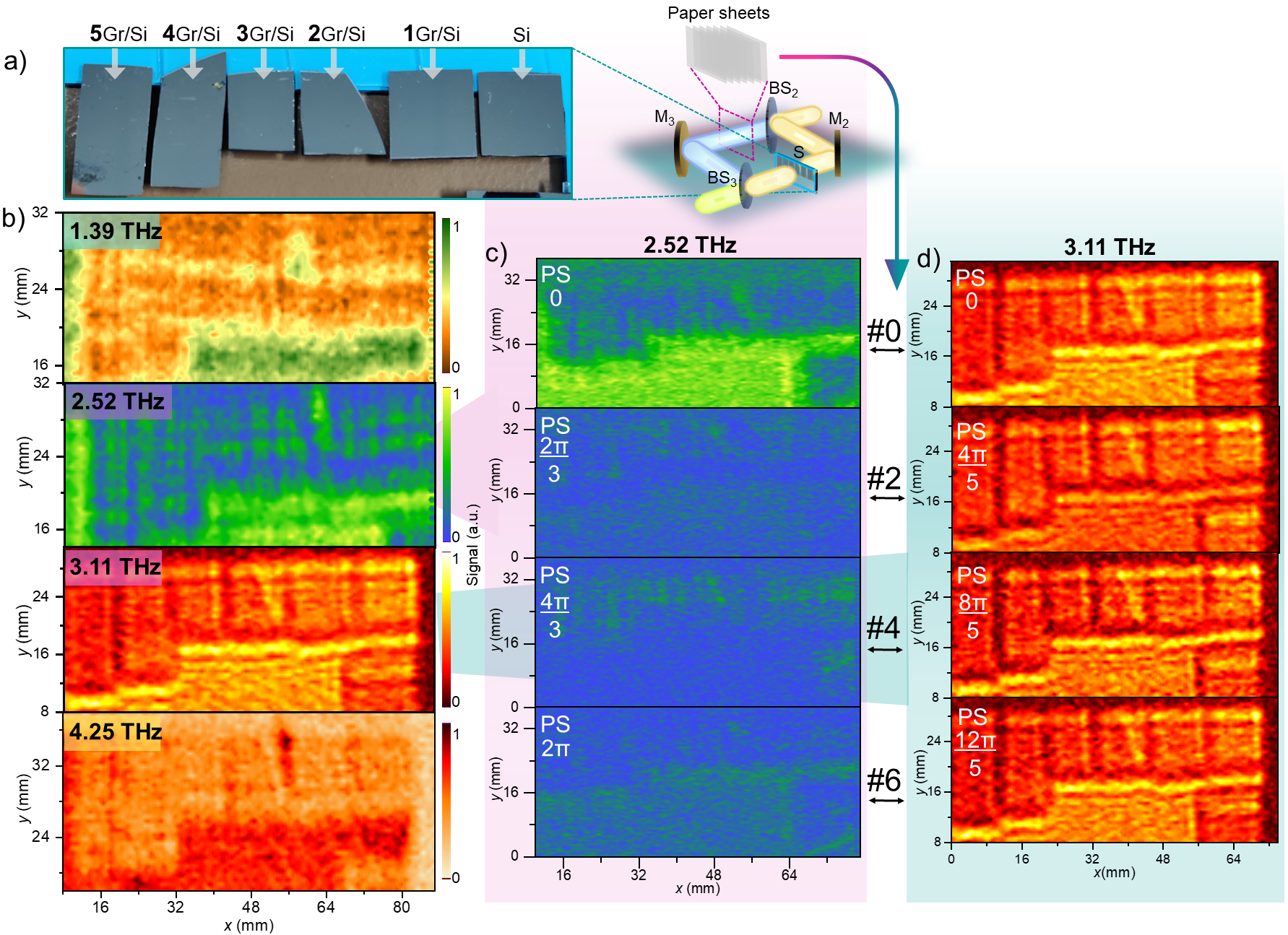}
    \caption{Holographic images of stacked graphene layers placed on silicon substrate. 
    {\bf Panel (a):} Photograph of the graphene samples where the different number of graphene layers (form 1 to 5) were transferred on high-resistivity silicon substrate. The inset on the right depicts part of the experimental setup, where S is the imaged sample, M$_2$ and M$_3$ are flat mirrors and BS$_2$, BS$_3$ are beam splitters. 
    {\bf Panel (b):} Experimentally obtained intensity distributions of the sample imaged at four different frequencies without introduction of the PS in the reference beam. Coloured scales are normalized to the maximum values. The contrast value for 1.39~THz is 0.51, for 2.52~THz -- 0.35, for 3.11~THz -- 0.83 and for 4.25~THz -- 0.66. 
    {\bf Panels (c-d):} Intensity distributions at 2.52~THz (c) and 3.11~THz (d) frequencies when the reference beam phase is shifted by placing 0, 2, 4 and 6 paper sheets. The resulting PS is indicated inside each image. Note that samples details are resolved at 3.11~THz precisely.}
    \label{fig:graphene}
\end{figure}

To illustrate a suitability of THz holograms for low-absorbing materials inspection, we performed a holographic study using samples of stacked graphene layers placed on the high resistivity silicon substrate as it is shown in Fig.~\ref{fig:graphene}(a). Thickness of the graphene monolayer is of only 0.3~nm, hence, its inspection in THz range is challenging due to low absorption is such thin material. Terahertz imaging together with simultaneous carrier optical modulation can be a rational way to test the material \cite{RusneJAP2022}, however, it requires additional optical excitation. Recently, it was demonstrated that THz structured light in compact THz imaging systems can serve as convenient tool for contactless inspection of stacked graphene layers \cite{RusneLSA2022}. Here, we perform further search for techniques that can be suitable for nondestructive inspection of 2D materials. In this study employing coloured digital THz holography, we used 5 samples with different number -- 1, 2, 3, 4 and 5 layers of graphene  -- transferred on a high resistivity silicon substrate. Silicon substrate itself (without graphene) served as a reference sample. Results are depicted in Fig.~\ref{fig:graphene}(b). As one can see, the optimal frequency for distinguishing numbers of graphene layers can be observed in Fig.~\ref{fig:graphene}(b), which corresponds to 3.11~THz. Images of the investigated samples when the PS using 2, 4 or 6 paper sheets was introduced to the reference beam are depicted in Fig.~\ref{fig:graphene}(c) for 2.52~THz and Fig.~\ref{fig:graphene}(d) for 3.11~THz frequencies. The latter exhibited better results in comparison to all others.

\begin{figure}[ht!]
    \centering
    \includegraphics[width=0.7\columnwidth]{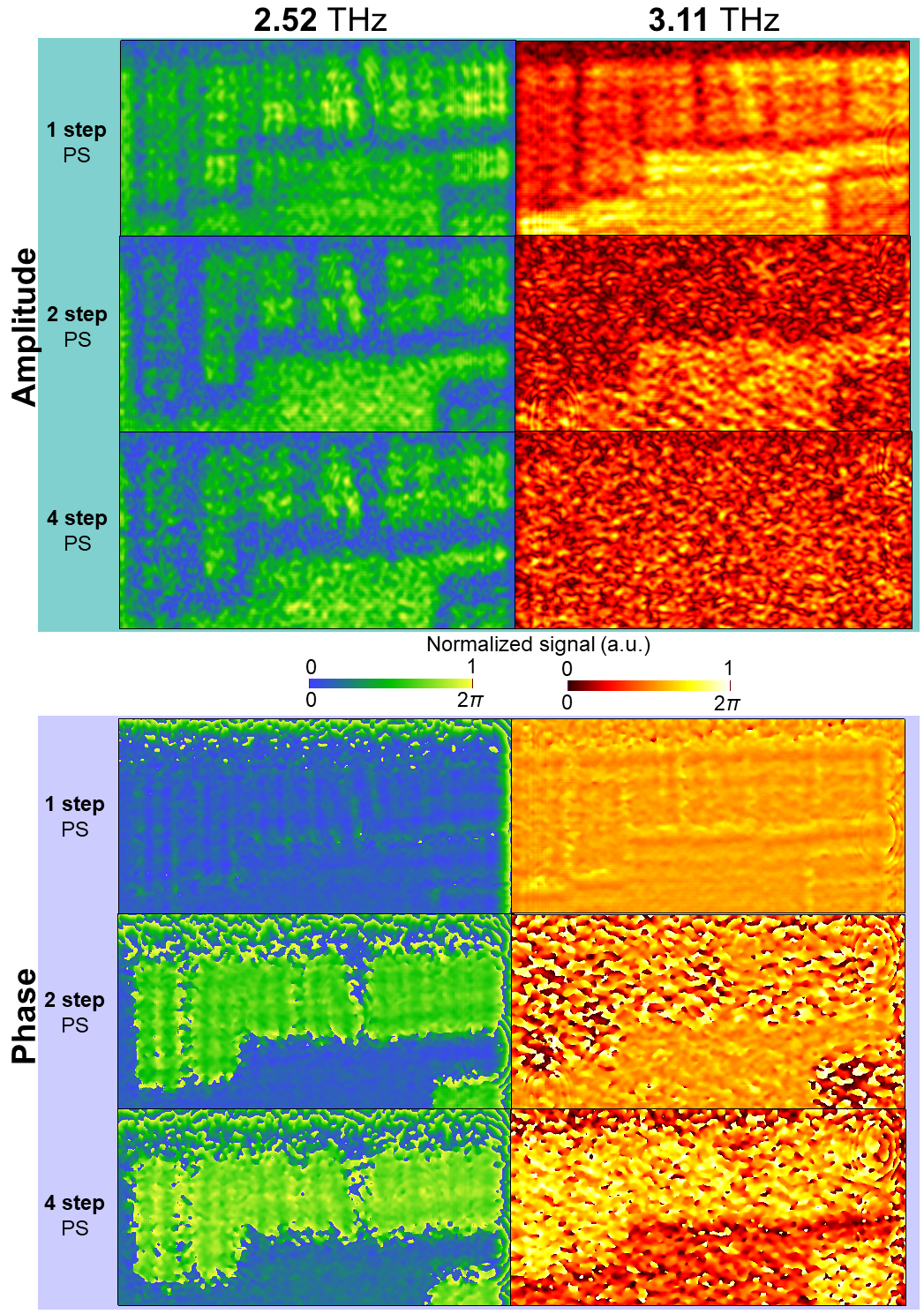}
    \caption{Intensity (top part) and phase (bottom part) distributions of the reconstructed holograms at four different frequencies: 1.39~THz, 2.52~THz, 3.11~THz and 4.25~THz. The imaged sample consists of different number of graphene layers (from 1 to 5) placed on the silicon substrate. In each case the reconstructions in the first row are obtained using a single experimentally measured hologram image. The reconstructions in the second row are obtained using 2-step PS method, while the reconstructions in the third row are obtained using 4-step PS method. Coloured scales are normalized to the maximum signal values. It should be underlined that using 4-step PS method a phase distribution corresponding to the phase delay map introduced by the sample is visible and recognisable in the phase distribution of the image reconstructed with the 4-step PS algorithm. At the same time the amplitude distribution remains at the constant level which stays in accordance with the assumptions and the experimental results.}
    \label{fig:graphene_reconstructions}
\end{figure}

For the graphene samples investigation the 4-step and 2-step phase-shifting algorithm was used for reconstruction, and, for comparison, the single hologram reconstructions were performed. 
Each hologram was scanned with 0.5~mm resolution to cover the total size of 18 x 80 mm. 
For the reconstruction, the whole scanned hologram area was placed inside 1024 x1024 pixel matrix which was rescaled 4~times to increase the reconstructed image resolution. The digital reconstruction was performed on 4096 x4096 pixels calculation matrix with sampling 0.125~mm. 
The registered intensity pattern was recalculated according to desired algorithm (4-step, 2-step phase-shifting or single hologram reconstructions) and propagated back at a distance of 95~mm corresponding to the distance between the object and the hologram planes. The amplitude and phase distributions in the image plane for hologram reconstructions for all frequencies and algorithms are shown in Fig. \ref{fig:graphene_reconstructions}.

As one can see, for frequency of 3.11~THz the reconstructed image amplitude distribution becomes more uniform for the PS methods with the increasing number of constituent holograms. It is clearly seen that the amplitude distribution for 4-step PS algorithm is uniform, while the phase distribution for the equivalent case delivers the most distinguishable areas which is straight forward related to the introduced PS of the transparent sample with varying thickness or having varying refractive index values. In Fig. \ref{fig:graphene_reconstructions} it can be seen that the amplitude distribution reconstructed from single hologram for the frequency of 3.11~THz is characterised by variation in intensity areas corresponding to the different amount of the graphene layers. Such result indicates that the graphene layers form an interferometric pattern which is capable of distinguishing the different number of layers from the amplitude distribution. However, this is the result of the interference of information coming out from the sample with additional noise distribution reconstructed from the same hologram distribution. 
It should be underlined that using 4-step PS method phase distribution corresponding to the phase delay map introduced by the sample is visible and recognisable in the phase distribution of the image reconstructed with the 4-step PS algorithm. We expect sample to be transparent with introduced particular PS and, thus, the only 4-step PS method gives the expected results without additional interferences. 
In case of the 2.52~THz frequency, the reconstructed amplitude distributions are much more uniform and the PS algorithms allow to reveal even small phase differences introduced by the sample. The more prominent is 4-step PS, which removes all the components resulting in additional noise and additional interferences in the reconstructed images.

To summarize, we demonstrated terahertz coloured digital holography ranging from 1.39~THz up to 4.25~THz. It is shown that it can be applied for the investigation of low-absorbing objects including objects consisting of very thin layers like graphene monolayers. In more detailed, inspection of stacked graphene layers  placed on high-resistivity silicon substrate is exposed using holographic recording based on an optically-pumped molecular THz laser operating at discrete emission lines of 1.39~THz, 2.52~THz, 3.11~THz, and 4.25~THz frequencies. It is revealed that phase-shifting methods  allow to qualitatively reconstruct coloured THz holograms with improved quality achieved by removing unwanted information related with so-called DC term and conjugated beam forming virtual image. Holographic reconstructions allow to observe the phase distributions introduced by the sample which is consistent with the theoretical assumptions about the sample.

\section*{Methods}

{\bf Holograms recording.}
 The experimental optical setup relying on the Mach-Zehnder interferometer was used for hologram recording. It is presented in Fig.~\ref{fig:setup}, where interfering beams are by yellow and blue colours. The additional violet beam was used as the reference beam for noise removal related to non-uniformities of the power emitted by the source.
 In order to use the uniform terms, we can define the following: the beam part in BS$_2$ -- M$_2$ -- BS$_3$ that goes through the sample is {\bf the object beam} -- marked as yellow; the beam part in BS$_2$ -- M$_3$ -- BS$_3$ passing through the PS is {\bf the reference beam} -- marked as blue, and the beam between BS$_1$ and D$_1$ is {\bf reference detector beam}-- denoted as violet.
 
The coherent THz radiation was generated by optically-pumped continuous wave molecular THz laser (FIRL100, Edinburgh Instruments,  marked as letter "E" in the setup scheme) delivering spectral lines at different frequencies. Emission lines at frequencies of 1.39~THz, 2.52~THz, 3.11~THz, and 4.25~THz using the power within 2--8~mW range aiming to keep laser modes stable were used for investigation. 

{\bf Holograms reconstruction.} 
The intensity pattern was recorded as the interference of two beams -- the object beam $U_{obj}$ and the reference beam $U_{ref}$ -- it formed a Fresnel hologram which can be numerically reconstructed. The intensity of the recorded by the detector hologram is given as:
\begin{equation}
I_{h}(x,y)=|U_{obj}(x,y)+U_{ref}(x,y)|^2,
\label{int-holo}
\end{equation}
which results in forming 4 terms. Two of them are not containing any phase information and form so-called DC term, one of them is corresponding to forming virtual image and the other one -- to real image. It is worth noting that only the last one is interesting for us in case of digital reconstruction -- because it forms a real image in the numerical plane corresponding to the object plane in real setup. Thus, all other terms are forming unwanted noise -- DC term propagating back and divergent light field distribution coming from virtual image. The light field distribution forming the real image can be extracted from all registered information using PS methods \cite{Yamaguchi1997}. Such methods assume recording different amount of successive holograms as interference of not changing object beam and reference beam undergoing particular PS between the following recording. Depending on the applied method a total or only partial removal of the unwanted terms can be obtained. In this study, a 4-step PS method (introducing 0, 0.5$\pi$, $\pi$ and 1.5$\pi$ phase shifts in the reference beam) is used as described in our previous investigation \cite{Siemion2021}. For comparison, a single hologram and a 2-step PS reconstruction, introducing 0 and $\pi$ PS in reference beam, were also conducted. The 4-step reconstructing algorithm utilizes for back propagation the intensity pattern given by the equation: 
\begin{equation}
I_{PS4}(x,y)=I_{h1}(x,y)-I_{h3}(x,y)+i[I_{h2}(x,y)-I_{h4}(x,y)],
\label{trans_PS4_simp}
\end{equation}
where $I_{h1}(x,y)$, $I_{h2}(x,y)$, $I_{h3}(x,y)$ and $I_{h4}(x,y)$ are the subsequently recorded holograms (intensity distribution being a result of interfering beams) with introduced PS of 0, 0.5$\pi$, $\pi$ and 1.5$\pi$ in the reference beam, respectively.
In the case of a 2-step PS algorithm, the algorithm simplifies to:
\begin{equation}
I_{PS2}(x,y)=I_{h1}(x,y)-I_{h3}(x,y).
\label{trans_PS4_simp}
\end{equation}
It should be taken into account that here the light field distribution forming a virtual image is introducing additional noise. In the case of single hologram reconstruction, only the backpropagation of the recorded intensity pattern forming a hologram is conducted.

\section*{Acknowledgements}

This research has received funding from the Research Council of Lithuania (LMTLT), agreement No [S-MIP-22-76].\\
Authors acknowledge Orteh Company for providing LS 6.0 software used here for numerical reconstruction of the recorded holograms, which is accessible in the Laboratory of Optical Information Processing at the Faculty of Physics in Warsaw University of Technology.\\
The data that support the findings of this study are available from the corresponding author upon reasonable request.

\section*{Author contributions statement}

A.S., L.M. and G.V. developed the concept of the study; L.M., I.G. and R.I.-P. performed experiments; Experimental data were analyzed by L.M., R.I.-P., I.G. and D.J.; A.S. performed reconstructions of obtained holography images; R.I.-P. fabricated graphene samples; K.I. and A.L. designed, fabricated and tested the CMOS THz sensor, the manuscript was written and edited by A.S., G.V., L.M., R.I.-P., I.G. and D.J.; All authors discussed and commented on this work.

\section*{Additional information}

The authors declare no competing interests.

\section*{Data availability}

The datasets used and/or analysed during the current study available from the corresponding author on reasonable request.

\bibliography{}

\end{document}